\documentclass[a4paper]{article}
\usepackage[dvipdfmx]{graphicx}
\usepackage{color}	
\usepackage[fleqn]{amsmath}
\usepackage{cases}
\usepackage{bm}

\title{Transient Behavior of Redox Flow Battery \\Connected to Circuit \\ Based on Global Phase Structure}

\author{Toko Mannari, Takashi Hikihara}

\newcommand{\figref}[1]{Fig.\,\ref{#1}}
\newcommand{\tabref}[1]{Table\,\ref{#1}}
\renewcommand{\eqref}[1]{Eq.\,(\ref{#1})}
\newcommand{\eqsref}[2]{Eqs.\,(\ref{#1}) and (\ref{#2})}

\newcommand{\eqshref}[2]{Eqs.\,(\ref{#1})--(\ref{#2})}
\newcommand{\cref}[1]{chapter \ref{#1}}
\newcommand{\mfigref}[2]{Fig.\,\ref{#1}(#2)}
\newcommand{\mfigsref}[3]{Figs.\,\ref{#1}(#2) and \ref{#1}(#3)}
\newcommand{\mfigshref}[3]{Figs.\,\ref{#1}(#2)--(#3)}
\newcommand{\ifigref}[1]{Figure.\,\ref{#1}}
\newcommand{\ifigsref}[1]{Figures.\,\ref{#1}}

\newcommand{\ieqshref}[2]{Equations.\,(\ref{#1})--(\ref{#2})}

\newcommand{\imfigref}[2]{Figure.\,\ref{#1}(#2)}

\newcommand{\imfigshref}[3]{Figures.\,\ref{#1}(#2)--(#3)}

\newcommand{\vII}{{\rm V^{2+}}}
\newcommand{\vIII}{{\rm V^{3+}}}
\newcommand{\vIV}{{\rm {VO^{2+}}}}
\newcommand{\vV}{{\rm {VO_2}^{+}}}
\newcommand{\e}{{\rm e^{-}}}
\renewcommand{\H}{{\rm H^{+}}}
\newcommand{\Cc}{c_{\rm c}}
\newcommand{\Ct}{c_{\rm t}}
\newcommand{\Cco}{c_{\rm c0}}
\newcommand{\Cmax}{{\rm c_{max}}}
\newcommand{\ac}{\alpha_\mathrm{c}}
\newcommand{\at}{\alpha_\mathrm{t}}
\newcommand{\Eeo}{\mathrm{E_\mathrm{e}^0}}

\newcommand{\Eec}{E_\mathrm{ec}}

\newcommand{\order}[1]{\times10^{#1}}
\newcommand{\M}{{\rm mol\,L^{-1}}}
\newcommand{\fr}{{\rm L\,min^{-1}}}

\newcommand{\odfrac}[2]{\frac{\mathrm{d}#1}{\mathrm{d}#2}}
\newcommand{\dd}[2]{\mathrm{d}#1/\mathrm{d}#2}
\newcommand{\oddfrac}[3]{\frac{\mathrm{d}^{#1}#2}{\mathrm{d}#3^#1}}
\newcommand{\NRNST}[1]{\Eeo\!+\!\frac{2RT}{F}\ln\frac{#1}{\Cmax-#1}}
\newcommand{\nrnst}[1]{1+\varepsilon\ln\frac{#1}{1-#1}}
\newcommand{\D}[1]{\Delta{#1}}
\renewcommand{\deg}{^{\circ}}

\begin{document}

\maketitle

\section{Introduction}
Power storage is the most important element of power grid to
overcome the restriction of simultaneous power balancing.
With an increase of installation of the renewable energy sources, 
power storage should inevitably absorb the fluctuation. 
Power storage has advantages as a large capacity and a quick response \cite{eng-science-storage-grid}. 
The large capacity is necessary for applications like load leveling,
and the quick response is necessary for applications like compensating sag  and smoothing of the output of renewable sources. 
Among the possible storage systems, 
a Redox Flow Battery (RFB) has a feature of scalability \cite{eng-Microgrids-RFB-characteristic,eng-EESRSGB-RFB}.
The feature is beneficial when installing the RFB in a various capacities of power grid. 
The feasibility of a Vanadium RFB for the load leveling was studied in Ohio \cite{eng-kazacos-review}. 
A Vanadium RFB was applied for the compensating of the momentum voltage sag in a semiconductor factory in Japan \cite{eng-SEI-RFB-renewable}.
A Vanadium RFB was tested for smoothing the output of the wind farm at Hokkaido, Japan \cite{eng-shigematsu-windfarm-shuntei,eng-SEI-RFB-renewable}.

An analysis on transient behaviors of an RFB is a key issue for the applications. 
Transient behaviors of the RFB deeply depend on the chemical reaction and the flow of the electrolyte. 
However, the dynamics of the transient behaviors contain mixed times scale transient dynamics. 
Moreover, if an RFB is connected to power grid, it will include the electric circuit restriction. 
And previous analysis of RFB is still far from the practical operation. 

An ODE model \cite{model-minghua-dynamical-simulation} is one of the more helpful tools to analyze the transient behavior. 
Researchers have developed several kinds of models 
such as empirical models, equivalent models, and ODE models to analyze transient behaviors of an RFB \cite{eng-model-review}. 
In \cite{model-minghua-dynamical-simulation},  an ODE model is proposed mainly based on chemical kinetics. 
This model represents change in a concentration of ions governing transient behaviors of the RFB. 
This model enables us to consider the chemical reaction, the flow of electrolyte, and the electric circuit restriction. 
The authors simulates electromotive force (EMF) at charging/discharging operation under constant current. 
The simulation results were confirmed by comparing to the experimental results in a constant current setting. 
The result showed that the model was valid. 

We use the model suggested in \cite{model-minghua-dynamical-simulation} 
and research transient behaviors of a Vanadium RFB in a response to a load variation, 
and consider the dynamical mechanism of the transient behaviors. 
In the research, 
the transient behaviors shows a non linearity. 
\section{Current at transient due to load variation}
This chapter discusses, 
transient behaviors of the RFB in a response to a load variation 
using the model based on chemical kinetics \cite{model-minghua-dynamical-simulation}. 
After introducing the model, 
the setting of simulation is explained. 
Here are found three types of transient behaviors in the simulations. 
\subsection{Model equation}
A model of the micro RFB was introduced based on chemical kinetics \cite{model-minghua-dynamical-simulation}. 
The model represents a change in concentration of ions and EMF. 
\tabref{tab:Nomenclature} shows the nomenclature.
\begin{table}[h]
  \centering
  \caption{Nomenclature.}\label{tab:Nomenclature}
   \begin{tabular}{ccc}\hline
    $\Cc$ & Concentration of ions ($\vII)$ in cell & $\M$ \\
    $\Ct$  & Concentration of ions ($\vII)$ in tank& $\M$\\
    $i$ & Current & $\mathrm{A}$\\
    $W$ & Flow rate & $\fr$\\
    $\ac$ & Volume of cell & $\mathrm{L}$ \\ 
    $\at$ & Volume of tank & $\mathrm{L}$ \\
    $F$ & Faraday constant & $\mathrm{C\,mol^{-1}}$ \\
    $\Cmax$ & Maximum of $\Cc$ & $\M$ \\
    $\Eeo$ & $\Eec$ at $\Cc=\Cmax/2$ & $\mathrm{V}$ \\\hline
   \end{tabular}
\end{table}

A change in concentration of ions is modeled by considering the reduction-oxidation reaction and the electrolyte flow. 
\ifigref{fig:structure} illustrates the scheme of the Vanadium RFB. 
The RFB mainly consists of a cell unit and a tank \cite{eng-EESRSGB-RFB}. 
In the cell, the chemical energy is converted to electrical energy by reduction-oxidation reaction. 
\begin{figure}[h]
 \centering
 \includegraphics[width=0.5\hsize]{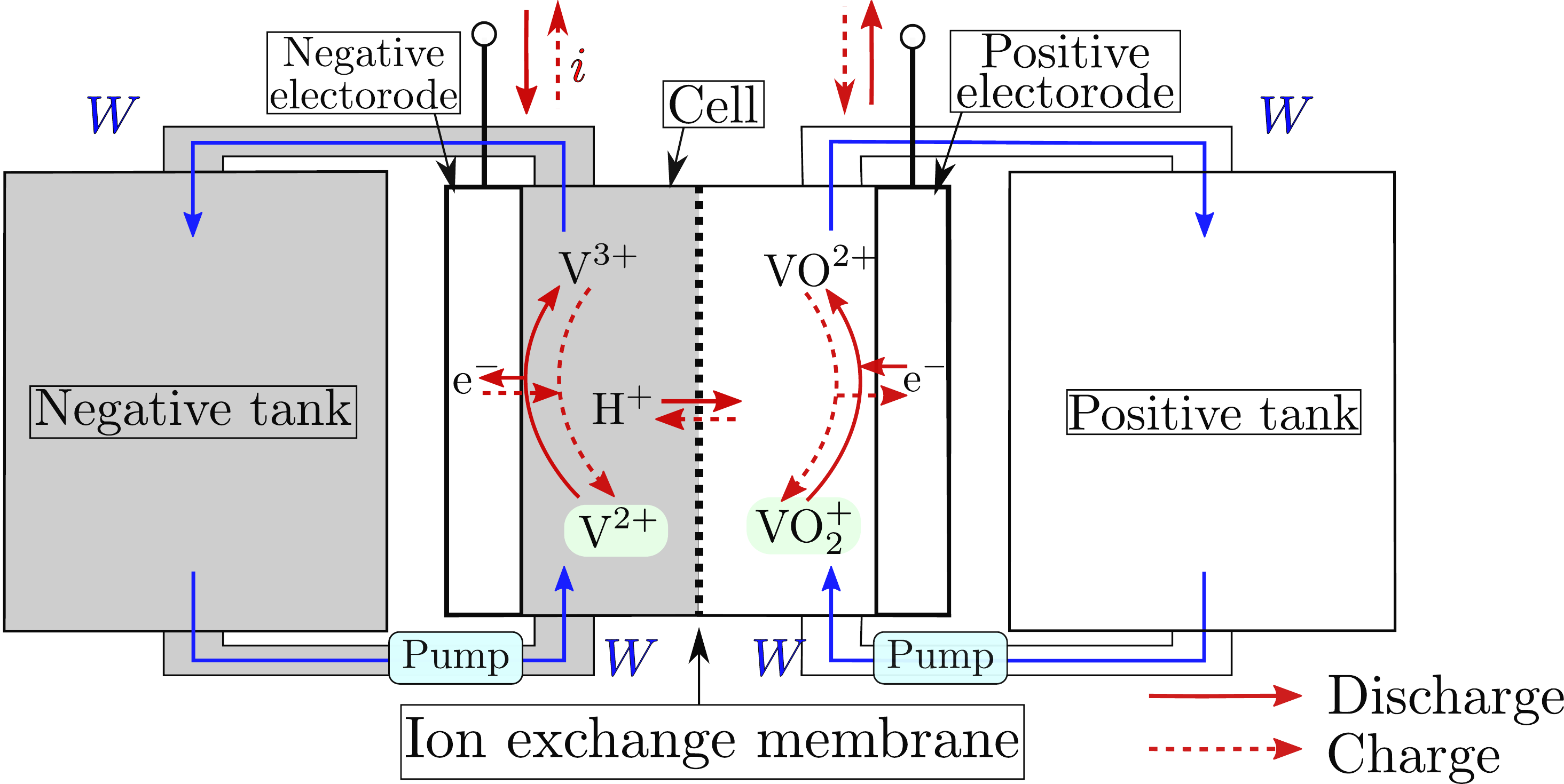}
 \caption{Scheme of Vanadium RFB.} \label{fig:structure}
\end{figure}
The reaction is represented as following \eqsref{eq:reaction_positive}{eq:reaction_negative}\cite{kazacos-characteristics-Vanadium-redox-flow-battery}. 
\begin{subnumcases}                               
 {}
  \vV + {\rm 2H}^+ + \e \rightleftharpoons \vIV + \mathrm{H_2O}\label{eq:reaction_positive}\\
  \vII \rightleftharpoons \vIII + \e\label{eq:reaction_negative}
\end{subnumcases}
The concentration of $\vV$ or $\vII$ at the cell governs the reaction. 
In the tank, the chemical energy is stored as a form of ion substantial. The electrolyte circulates between the cell and the tanks by pumps. 
The circulation makes a flow in the cell to supply the concentration of ions
for the continuous reaction described by \eqsref{eq:reaction_positive}{eq:reaction_negative}. 
Here we set following assumptions, 
\begin{itemize}
 \item Chemical kinetics is restricted by the current in the external circuit. 
 \item The electrolyte keeps the consistency of ions.  
 \item The time delay of circulation of electrolyte is negligible. 
 \item The concentration of ions ($\vII$) in the negative half cell is equal to the concentration of ions ($\vIV$) in the positive half cell. 
\end{itemize}
\begin{eqnarray}
 \oddfrac{2}{\Cc}{t}\!&=&\!-W\left(\frac{1}{\ac}\!+\!\frac{1}{\at}\right)\!\odfrac{\Cc}{t}
 \!-\!\frac{Wi}{\ac\at F}\!-\!\frac{1}{\ac F}\odfrac{i}{t}\label{eq:dCc}\\
 \odfrac{\Ct}{t}\!&=&\!-\frac{\ac}{\at}\odfrac{\Cc}{t}-\frac{i}{\at F} \label{eq:dCt}
\end{eqnarray}
When the RFB is charging/discharging, $i$ is positive/negative respectively. 

The model of the EMF is derived based on Nersnst's equation \cite{nernst-equation-eng}. 
Nersnst's equation gives the equilibrium potential of the battery. 
And from an experiments in \cite{kazacos-mechanism-water-transfer}, 
the change in the concentration of $\H$ is also small enough to be ignored.
Then, the model of the EMF is derived as following. 
\begin{equation}
 \Eec=\NRNST{\Cc}\label{eq:Eec}
\end{equation}
\subsection{Simulation setting}\label{setting}
Here explains the setting of the simulation. 
\ifigref{fig:slide_circuit} shows the target system of the simulation.
\begin{figure}[h]
 \centering
 \includegraphics[width=0.5\hsize]{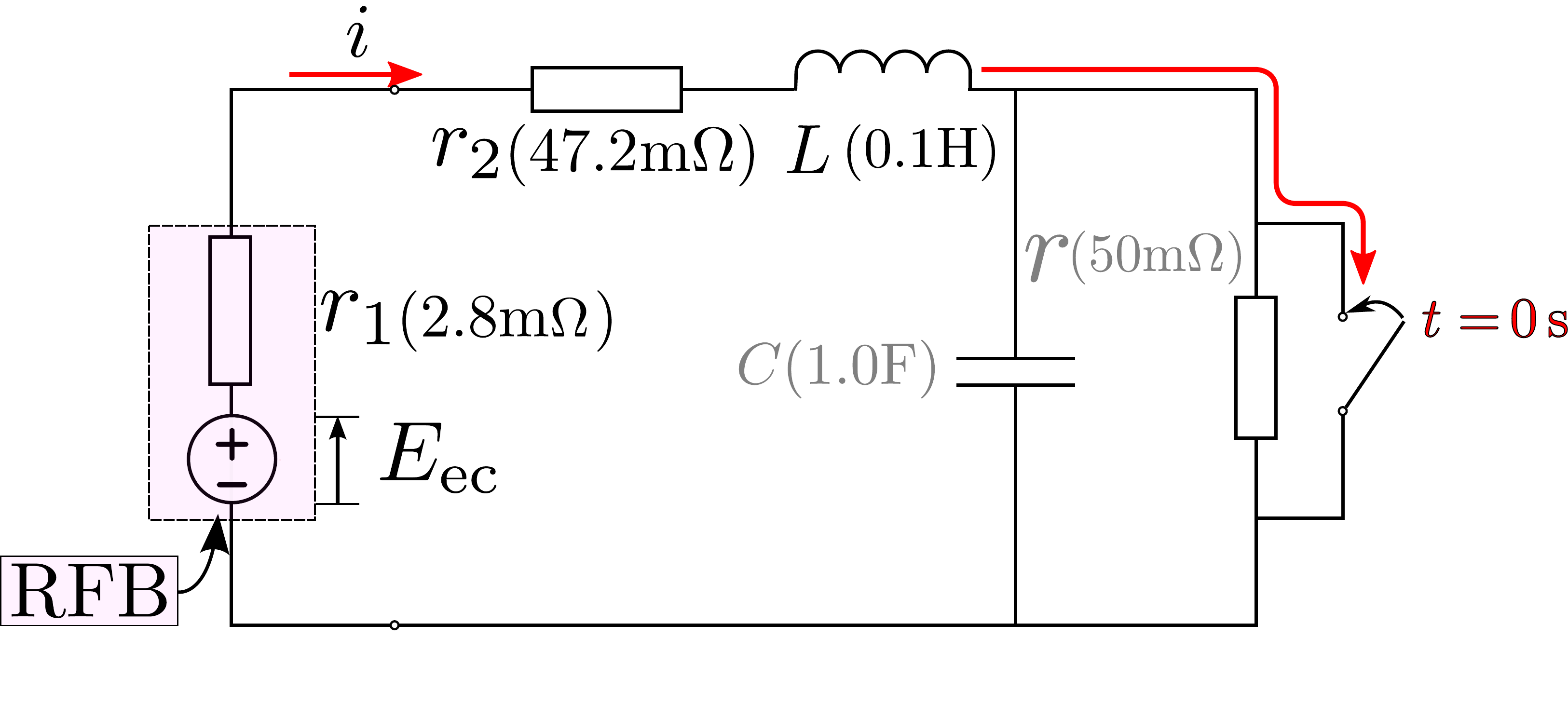}
 \caption{Target system of simulation which include RFB. The load of RFB has changed in a step at $t=0\,\mathrm{s}$.}
 \label{fig:slide_circuit}
\end{figure}
At $t=0\,\mathrm{s}$, the switch turns on and the load is forced to a step change. 
The setting enables us to study the response of the RFB to the load variation. 
From KVL, following equation is obtained. 
\begin{equation}
 \odfrac{i}{t}\!=\!-\frac{1}{L}\!\left\{\!(r_1+r_2)i\!-\!\left(\!\NRNST{\Cc}\!\right)\!\right\}\label{eq:simulation_di}
\end{equation}
By using \eqsref{eq:dCc}{eq:simulation_di}, the following equation is derived.
\begin{equation}
 \oddfrac{2}{\Cc}{t}\!=\!-W\!\left(\!\frac{1}{\ac}\!+\!\frac{1}{\at}\right)\!\odfrac{\Cc}{t}
 +\!\left(\frac{r_1+r_2}{L}-\frac{W}{\at}\!\right)\!\frac{i}{\ac F}
 -\frac{1}{\ac FL}\left(\NRNST{\Cc}\right)\label{eq:simulation_dCc}
\end{equation}
\ieqshref{eq:Eec}{eq:simulation_dCc} enable to simulate the transient behaviors of the RFB in this setting. 
\tabref{tab:parameter} shows the parameters of the RFB. 
\begin{table}[h]
 \begin{center}
  \caption{Parameters of RFB connected to circuit (ref.\cite{model-minghua-dynamical-simulation}).}
  \begin{tabular}{ccc}\hline
   Parameter&Value&Unit\\\hline
   $\ac$&$0.100$&$\mathrm{L}$\\
   $\at$&$0.900$&$\mathrm{L}$\\
   $T $&$307$&${\rm K}$\\
   $\Cmax$&$1.70$&$\mathrm{mol\,L^{-1}}$\\\hline
   \label{tab:parameter}
  \end{tabular}
 \end{center}
\end{table}
The initial condition of the simulation is determined by putting assumptions as followings. 
\begin{itemize}
 \item $\dd{i}{t}=0$ $(t\rightarrow-0)$. 
 \item $\Cc=\Ct$ $(t\rightarrow-0)$. 
\end{itemize}
In this simulation, forth-order Runge-Kutta method \cite{Numerical-Recipes-in-C} is adopted to solve the ODE. 
The step size is fixed at $h=0.001\,\mathrm{s}$. 

\subsection{Result of simulation}
In this section, first, the results of the simulation is explained and 
the transient behaviors are grouped into three types. 
Second, the consumption of the ions in the tank is examined in rate with comparing the transient behaviors of current. 

The simulation reveals that 
the current depends on the flow rate 
and behaves by different mechanics. 
\ifigref{fig:simulation} shows the result of the simulation.
\begin{figure}[h]
  \centering
 \includegraphics[width=1.0\hsize]{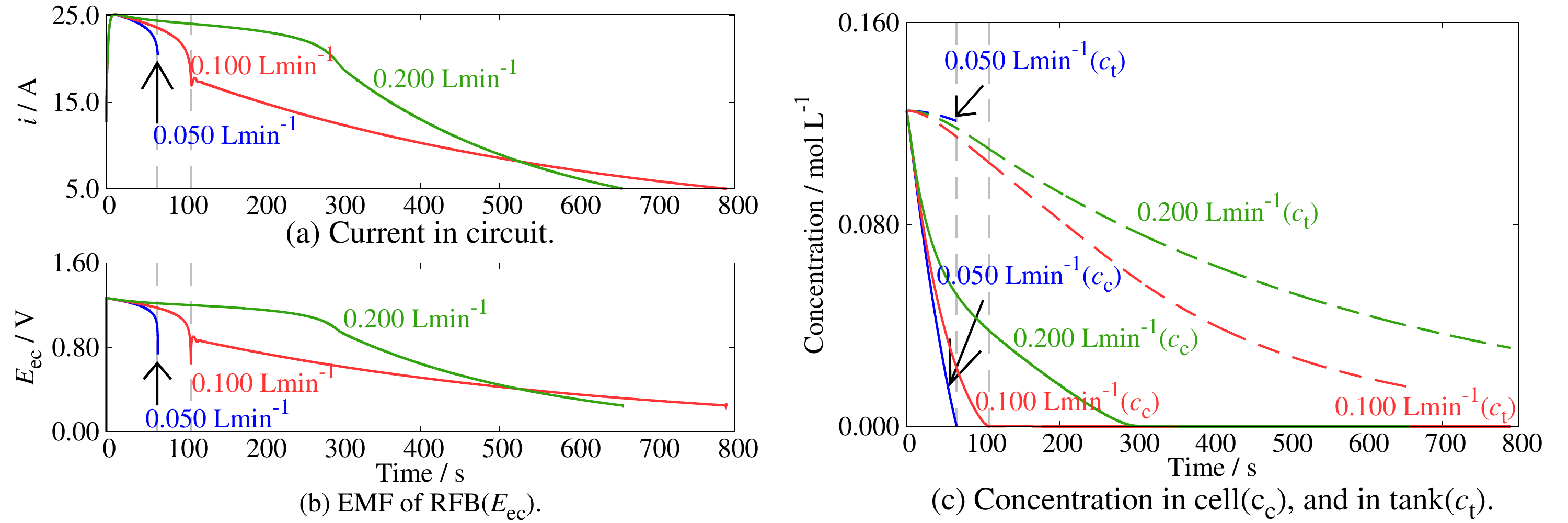}
 \caption{Simulated results of transient behaviors of RFB due to load variation. The initial value of the concentration of ions in the cell ($\Cco$) is fixed at $0.125\,\M$. The flow rate is set at $W=0.050\,\fr$, $0.100\,\fr$, and $0.200\,\fr$. The transient behaviors are categorized into three types (Case1--3). }
 \label{fig:simulation}
 \end{figure}
The initial value of the concentration of ions in the cell $\Cco$ is fixed at $0.125\,\M$. 
And the flow rate is set at $W=0.050\,\fr$, $0.100\,\fr$, and $0.200\,\fr$. 
As \mfigsref{fig:simulation}{a}{b} show, 
at $W=0.050\,\fr$, the current and the EMF show
sharp drops and 
the discharging simultaneously stops at the drops. 
We call the mode Case1. 
On the other hand, at $W=0.200\,\fr$, 
the current and the EMF show 
gradual decreases 
and discharges slowly. 
The discharging times is long. 
We call the mode Case2. 
At $W=0.100\,\fr$,
the current and the EMF show an oscillation. 
It is called Case3. 
The transient behaviors due to other initial conditions 
are also categorized to the Case1, 2, and 3. 
There seems to be critical states in \figref{fig:simulation}.

The difference of the transient behaviors among the cases 
is corresponds to the consumption rate of ions in the tank. 
As showed in \mfigref{fig:simulation}{c}, 
almost all the ions in the tank 
are not consumed at $W=0.050\,$. 
It suggests that the transient behaviors 
are allocated to the consumption rate of the ion in the tank.

In order to confirm the assumption, the consumption rate of the ions in the tank is calculated for $W=0.001\,\fr$--$0.200\,\fr$ and $\Cco=0.01\,\M$--$1.00\,\M$. 
The consumption rate is defined as following. 
\begin{equation}
 \varepsilon_\mathrm{t}=\frac{c_\mathrm{t0}-c_\mathrm{tf}}{c_\mathrm{t0}}
\end{equation}
Here, $c_\mathrm{tf}$ denotes the concentration of ions in the tank at the time when the discharging has finished. 
\begin{figure*}[h]
 \centering
 \includegraphics[width=1.0\hsize]{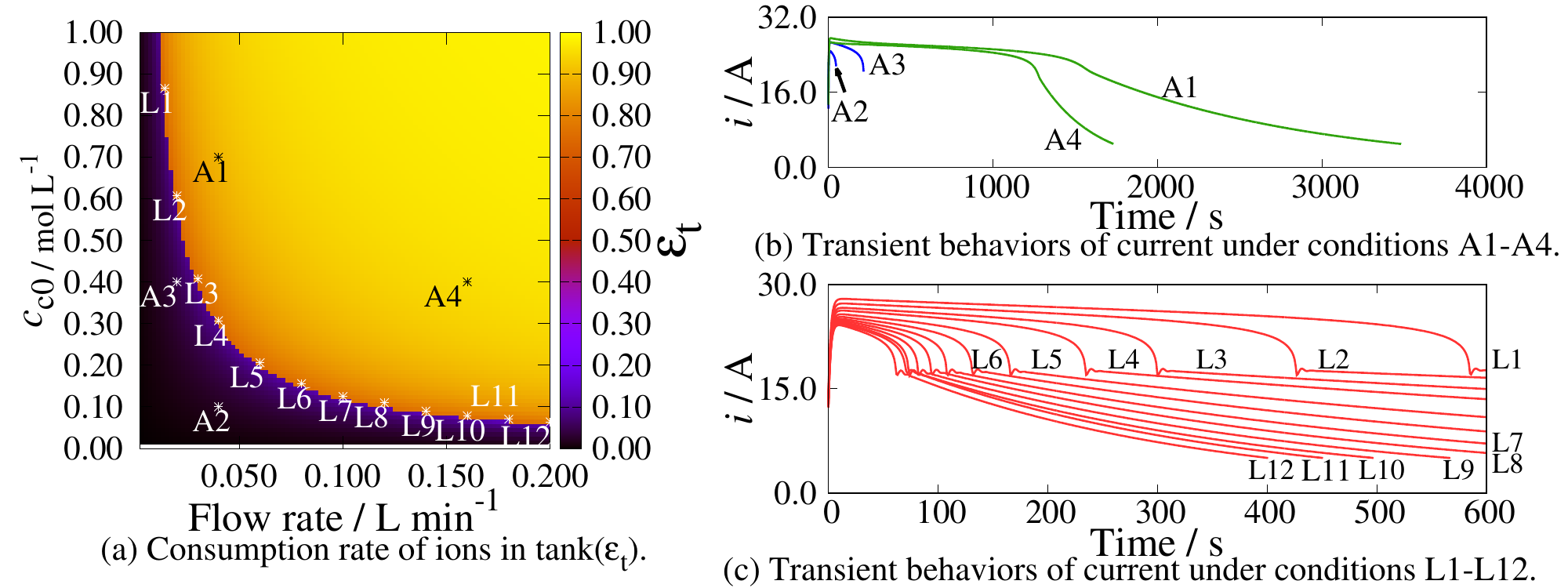}
 \caption{Relationship between transient behaviors of current and consumption rate of ions in tank. (a)Consumption rate of the ions in the tank. A line divides the graph into two areas. (b)Transient behaviors of current simulated under conditions A1 and A4 (above the line) and A2 and A3 (below the line). When conditions are set at A1 and A4, the transient behaviors of current are categorized into Case2, and when conditions are set at A2 and A3, the transient behaviors of current are categorized into Case1. (c)Transient behaviors of current simulated under conditions (on the line). When conditions are set at L1--12, the transient behaviors of current are categorized into Case3. }
 \label{fig:mapdate}
\end{figure*}
The results are showed in \mfigref{fig:mapdate}{a}. 
As showed in the figure, a line divides the graph into two regions. 
The region above the line represents 
the complete consumption of ions in the tank, 
and the below represents 
the remaining of ions until the end of discharging. 
Transient behaviors of current are simulated 
under conditions A1--4 as showed in \mfigref{fig:mapdate}{b}.  
When the conditions are set at A1 and A4, 
the current behaves as classified in Case2. 
When the conditions are set at A2 and A3, 
the current behaves as classified in Case1.
Transient behaviors of current are also simulated
under conditions L1--12 as showed in \mfigref{fig:mapdate}{c}. 
When the conditions are set at L1--12, 
the current behaves as classified in Case3. 
\ifigref{fig:mapdate} shows that 
the transient behaviors 
are actually correspond to the consumption of the ions in the tank. 
\section{Dynamical mechanism of transient behaviors}
There seems to exist a mechanism which governs the transient behaviors to lead them into three types.
In this chapter, we are going to focus on a dynamical mechanism of the transient behaviors. 
A dynamical mechanism of the three types of the transient behaviors is discussed with a dimensionless model in this chapter. 
The following discussions are from the view point of local dynamics around fixed point and global phase structures. 
\subsection{Analysis around a fixed point}
The target system is modeled by following \eqshref{eq:non_dimension_x1}{eq:non_dimension_x3}. 
\begin{subnumcases} 
\displaystyle
\odfrac{x_1}{\tau}\!=\!x_2\label{eq:non_dimension_x1}\\
\displaystyle
\odfrac{x_2}{\tau}\!=\!-\beta x_2\!+\!(1\!-\!\gamma)x_3\!-\!\left(\nrnst{x_1}\right)\label{eq:non_dimension_x2}\\ 
\displaystyle
\delta \odfrac{x_3}{\tau}\!=\!-\!x_3\!+\!\nrnst{x_1}\label{eq:non_dimension_x3}
\end{subnumcases}
Where $x_1$, $x_2$, and $x_3$ are the dimensionless variables 
corresponding to $\Cc$, $\dd{\Cc}{t}$, and $i$ with the constants $\hat{c},\,\hat{t}$, and $\hat{i}$. 
The constants $\hat{c},\,\hat{t}$, and $\hat{i}$ are given by \eqref{eq:constants}.
\begin{equation}
 \hat{c}=\Cmax,\,\hat{t}=\sqrt{\frac{\ac FL\Cmax}{\Eeo}},\,\hat{i}=\frac{\Eeo}{r_1+r_2}\label{eq:constants}
\end{equation}
Where the parameters $\beta,\,\gamma$, $\delta$, and $\varepsilon$ are given by \eqref{eq:parameter}. 
\begin{equation}
 \beta=W\left(\frac{1}{\ac}+\frac{1}{\at}\right)\hat{t},\,
 \gamma=\frac{WL}{\at(r_1+r_2)},\,
 \hat{t}'=\frac{L}{(r_1+r_2)},\,
 \delta=\frac{\hat{t}'}{\hat{t}},\,
  \varepsilon=\frac{2RT}{F\Eeo}
 \label{eq:parameter}
\end{equation}
From $\dd{{\bm x}}{\tau}=0$ (${\bm x}=[x_1\,x_2\,x_3]^T$), 
the system has only one fixed point ${\bm x}^{\ast}=[3.51\order{-12}\,0\,0]^T$. 

The local stability is analyzed with eigen values of a linearlized model of \eqshref{eq:non_dimension_x1}{eq:non_dimension_x3} \cite{wiggins-introduction-eigen-value}. 
The linearization around ${\bm x^*}$ is described by \eqref{eq:linearization}. 
\begin{equation}
 \D {\bm x}=A \D{\bm x}\label{eq:linearization}
\end{equation}
Here, $\D {\bm x}$ is ${\bm x}-{\bm x^{\ast}}$. 
The matrix associated with the linearized model is given as following \eqref{eq:A}. 
\begin{equation}
 A=\left[
    \begin{array}{ccc}
     0 & 1 & 0\\
     -f({\bm x^{\ast}}) & -\beta & 1-\gamma\\
     f({\bm x^{\ast}})/\delta & 0 & -1/\delta
    \end{array}
   \right] \label{eq:A}
\end{equation}
Where $f({\bm x})$ is given by \eqref{eq:f}. 
\begin{equation}
 f({\bm x})=\frac{\varepsilon}{x_1(1-x_1)}\label{eq:f}
\end{equation}
\tabref{tab:eigen_value} shows eigen values of $A$ for $W=0.050\,\fr$, $0.100\,\fr$, and $0.200\,\fr$.
\begin{table}[h]
 \begin{center}
  \caption{Eigen values of $A$ for $W=0.050\,\fr$, $0.100\,\fr$, and $\,0.200\,\fr$.}
  \begin{tabular}{|c|c|}\hline\label{tab:eigen_value}
   $W(\fr)$&Eigen value of $A$\\\hline
   &$-8.70+1.49\order{5}\mathrm{j}$\\
   $0.050$&$-8.70-1.49\times10^5\mathrm{j}$\\
   &$-3.17\times10^{-2}$\\\hline
   &$-8.84+1.49\order{5}\mathrm{j}$\\
   $0.100$&$-8.84-1.49\times10^5\mathrm{j}$\\
   &$-6.34\times10^{-2}$\\\hline
   &$-9.13+1.49\order{5}\mathrm{j}$\\
   $0.200$&$-9.13-1.49\times10^5\mathrm{j}$\\
   &$-12.7\times10^{-2}$\\ \hline
  \end{tabular}
 \end{center}
\end{table}
For these three parameters, the real part of eigen values are negative. 
Then, the fixed point ${\bm x}^{\ast}$ is locally stable, for these $W$. 
The stability implies the convergence of the solutions in the neighborhoods of the fixed point. 

On the other hand, the dynamics apart from the fixed point is governed by the vector flow of global phase structure. 
\ifigref{fig:vector_renormalize} illustrates the trajectory solutions categorized Case1--3.
\begin{figure*}[h]
 \includegraphics[width=1\hsize]{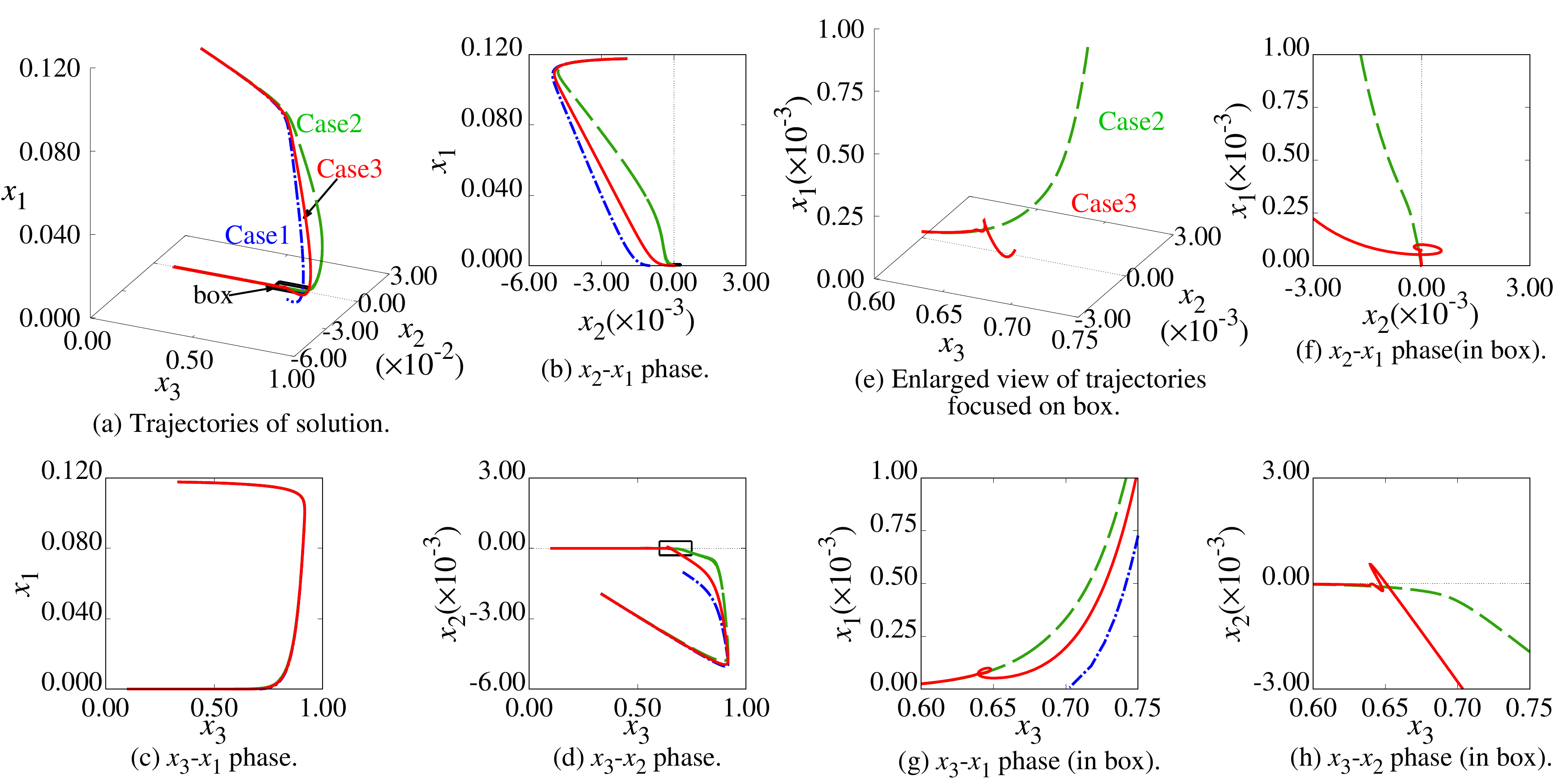}
 \caption{Trajectory solutions classified as Case1--3. (a)--(d)The trajectory solutions classed as Case1 clearly does not converge to the fixed point $\bm{x}^{\ast}=[3.51\order{-12}\,0\,0]^T$. (e)--(h)Enlarged view of the trajectories. The trajectory solution categorized into Case3 does not oscillate around the fixed point. }
 \label{fig:vector_renormalize}
\end{figure*}
\imfigshref{fig:vector_renormalize}{a}{d} show trajectories of the solutions categorized into Case1.
The trajectories will not converge to the fixed point. 
\imfigshref{fig:vector_renormalize}{e}{h} show the trajectories of the solutions categorized into Case3.
They will not oscillate around the fixed point.
The obtained solutions show the features of the global phase structure which is governed by the non linearity. 
\subsection{Global phase structure}
This section discusses 
a part of the global phase structure of the solutions and the vector field. 
The global phase structure is 
magnified at small scale 
for visualizing the vector field. 
The approach is based on the knowledge of fast-slow system 
\cite{hoshino-sent-multiple-time-scale-dynamics-chapter1}. 

\begin{figure*}[h]
 \centering
 \includegraphics[width=1.0\hsize]{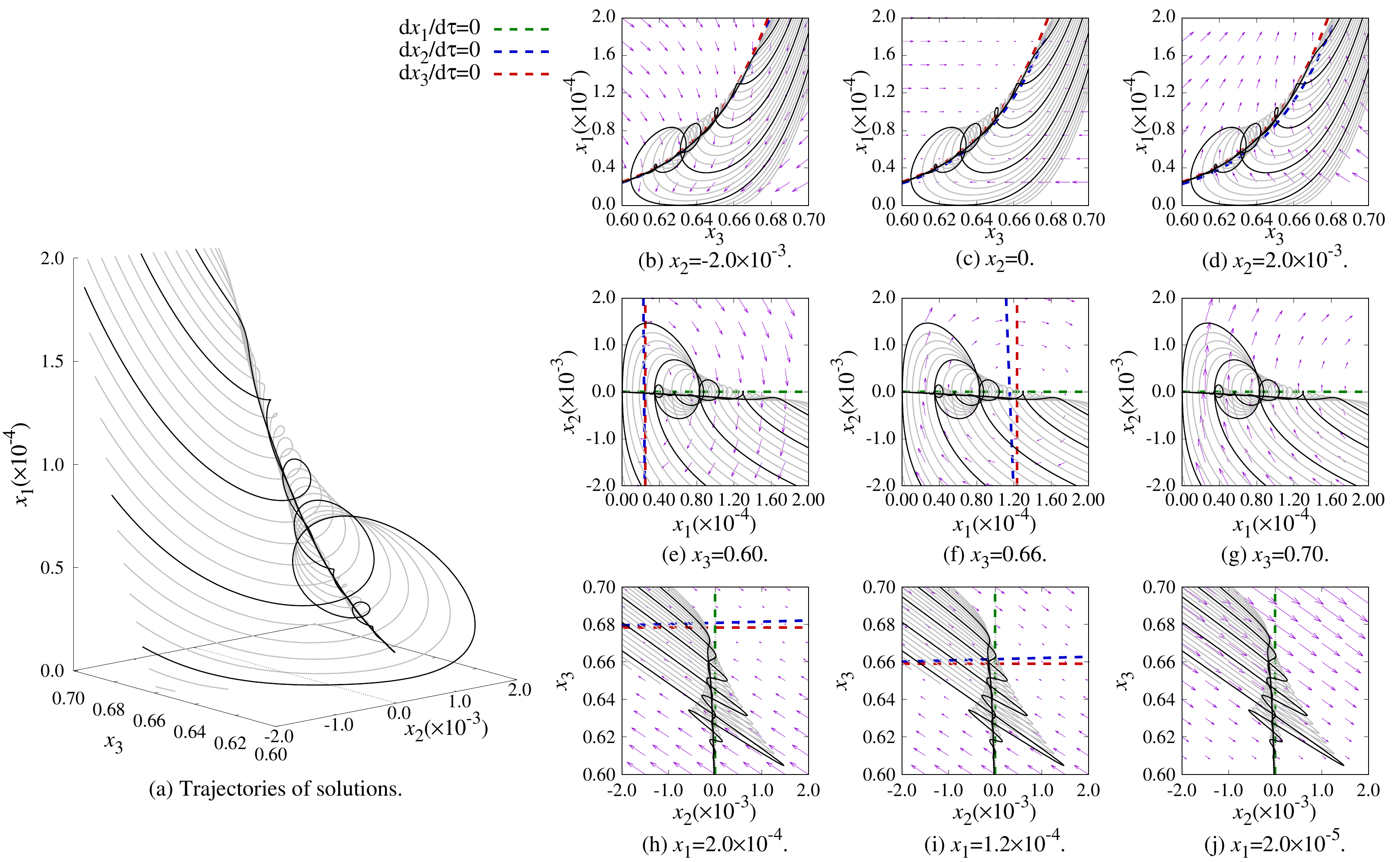}
 \caption{Global phase structure of solutions and vector field at $W=0.100\,\fr$. The area where the oscillation occur is focused on. In this area, the concentration of ions ($x_1$) is low and the current ($x_3$) changes largely compared to other two variables. (b)--(j)Vector field and nulkline is showed. Fast flow transverses to the nulkline $\dd{x_3}{\tau}=0$. }
 \label{fig:vector}
\end{figure*}
Figure \ref{fig:vector}(a) shows a part of the solutions for $W=0.100\,\fr$. 
In this figure, the oscillations are found 
in a small scale which is $x\sim 10^{-4}$. 
$x_1$ corresponds to $\Cc$. 
Then, oscillations occur at low concentration of ions at the cell. 
On the other hand, the scale of changes in $x_3$, 
corresponding to $i$, 
is much bigger than both $x_1$ and $x_2$. 
This implies the current is sensitive to the change in concentration of ions. 
The scale reveals 
the magnification characteristics 
in the battery from ion density to the current. 

The vector field is showed in \mfigshref{fig:vector}{b}{j}. Figures \ref{fig:vector}(b)--(d) show 
the vector field 
and the trajectories of the solutions 
on the planes $x_2$=$-2.0\order{-3}$, $0$, and $2.0\,\order{-3}$. 
In the figures, 
the trajectories converge to the nulkline $\dd{x_3}{\tau}$=$0$. 
When $\dd{x_3}{\tau}$=$0$ is established,
the RFB is limited by circuit condition. 
Fast flows appear along $x_3$,
due to the value of $\delta$ in \eqshref{eq:non_dimension_x1}{eq:non_dimension_x3}. 
The value of $\delta$ implies the ratio of time scale of current to that of change in concentration of ions. 
\ifigsref{fig:vector}(e)--(g) show the vector field and the trajectories of the solutions 
on the planes $x_3$=$0.60$, $0.66$, and $0.70$. 
As the figures show, the trajectories converge to the nulkline $\dd{x_1}{\tau}=0$.
\ifigsref{fig:vector}(h)--(j) show the vector field and the trajectories of the solutions 
on the planes $x_1$=$0.50\order{-4}$, $1.0\order{-4}$, and $2.0\order{-4}$. 
As the figures show, the trajectories converge to the nulkline $\dd{x_1}{\tau}=0$.
From the vector field,
it is found that
the system is fast-slow system.
But the system can not be divided into slow-subsystem and fast-subsystem clearly. 

The mechanism of the oscillations can be discussed by linearization with taking a plane perpendicular to a slow variable.
$x_1$ is one of the slow variables. 
The values of $x_1$ are swapped in $f({\bm x})$ from $x_1$=0.0000010 to $x_1=$0.005. 
 And the eigen values on each planes are shown in \mfigsref{fig:eigen_value}{a}{b}.
 In the figures, 
 all of the eigen values are located
 on complex plane with $\mathrm{Re[\lambda_i]}$$<$0\,$(i=1,\,2,\,3)$. 
 Then, a bifurcation appears at $x_1$=$5.72\order{-4}$. 
 As shown in $x_1>5.72\order{-4}$, 
 the eigen values become real. 
 It implies that 
 there does not appear any oscillatory transient behaviors in $x_1>5.72\order{-4}$. 
 And $\lambda_1$ and $\lambda_2$ come close to each other as $x_1$ decreases. 
 At $x_1=5.72\order{-4}$ they conjoint. 
 After the conjoint by $x_1$ decreasing, 
 for $x_1<5.72\order{-4}$, 
 $\lambda_1$ and $\lambda_2$ apart with imaginary parts, 
 and the values increase as $x_1$ decreases. 
 This clearly shows the appearance of oscillation in $x_1<5.72\order{-4}$. 
 \imfigref{fig:eigen_value}{c} shows the 
 phase lags of $\lambda_i$\,($i=1,\,2,\,3$). 
 $x_1\to 0$, the lags converge to $90\deg$, $90\deg$, and $180\deg$.
 Then, the phase lags of the oscillations are also expected to converge $90\deg$, $90\deg$, and $180\deg$.
 Actually, in \mfigshref{fig:vector_renormalize}{b}{j}, 
 the oscillation in $x_1$ has about $90\deg$ phase lag to $x_2$ and $x_3$, 
 and the oscillation in $x_2$ has about $180\deg$ phase lag to $x_3$. 
 The eigen values in \figref{fig:eigen_value} certainly shows the mechanism of the oscillation. 
 \begin{figure}[h]
 \centering
 \includegraphics[width=1.0\hsize]{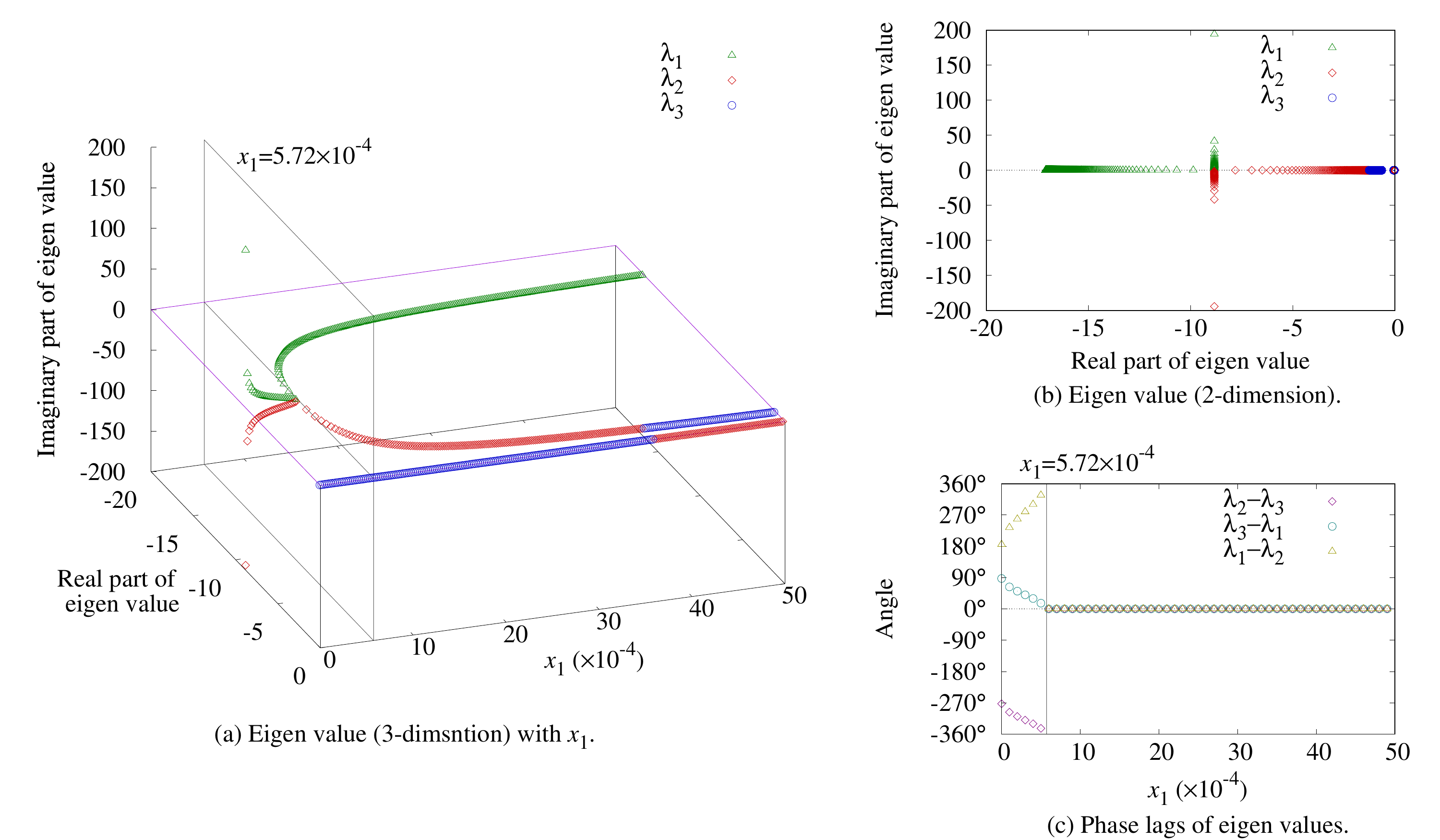}
  \caption{(a)(b)Eigen values of Jacobin matrix $A$. A bifurcation appears at $x_1$=$5.72\order{-4}$.
  In $x_1<5.72\order{-4}$, appearance of oscillations is showed by imaginary part of eigen values.
  (c)Phase lags of eigen values $\lambda_i$\,($i=1,\,2,\,3$).
  Phase lags of oscillations are expected to converge to $90\deg$, $90\deg$, and $180\deg$.}
 \label{fig:eigen_value}
\end{figure}
\section{Conclusion}
We discussed the transient behaviors of the RFB in a response to a load variation, 
and we found the dynamical mechanism of the transient behaviors. 
Results are summarized as followings. 
\begin{enumerate}
 \item Three types of transient behaviors appear and one of them causes oscillations.
       The transient behaviors depend on the value of flow rate and the initial value of the concentration of ions. 
 \item The global phase structure determines the transient behaviors.
       From the global phase structure, it is clarified that the behaviors are limited by electrical circuit restriction
       due to the value of a parameter which corresponding to time scale ratio of current to the changing ratio in concentration of ions. 
       It is also clarified that a necessary condition for occurrence of the oscillations are governed by the low concentration of ions. 
\end{enumerate}

The results show that
several kinds of phenomena appear in transient behaviors of single RFB, 
when the RFB is connected to power grid. 
According to flow rates or initial values of concentration of ions, 
the discharging of the RFB stops before it consumes the all of ions. 
On the other hand, there are conditions which arise oscillations. 
In the worst case, it may cause resonances with external systems. 
To avoid these anomalous phenomena, 
a control method of multi layer must be developed with considering chemical, fluid dynamics,
and electrical circuit restriction. 

The transient behaviors in the wide range are based on the global dynamics,
so that the control method must pay attention to the mechanism. 
However, the whole dynamical structure has not been figured out through the simulations. 
This paper brought us an interesting dynamical behavior with a difficulty of a mixed time scale system
because of the global phase structure along the slowest variable. 

\end{document}